\documentclass[prb,a4paper,showpacs,showkeys]{revtex4}

\usepackage{latexsym}
\usepackage{amsmath}
\usepackage{dcolumn}

\usepackage{color}

\usepackage[dvips]{graphicx}
\graphicspath{{./figs/}}

\usepackage{subfigure}

\newcommand {\qref}[1]{Ref.~\onlinecite{#1}}

\newcommand {\qfig}[1]{Fig.~\ref{#1}}
\newcommand {\queq}[1]{(\ref{#1})}
\newcommand {\qeq}[1]{Eq.~\queq{#1}}

\newcommand {\Er}{E_r}
\newcommand {\eVA}{eV/\AA$^3$}

\newcommand {\eps}{\epsilon}
\newcommand {\gamu}{\gamma_u}
\newcommand {\gams}{\gamma_s}

\newcommand {\es}{elastic stiffness}
\newcommand {\SF}{stacking fault}
\newcommand {\two}{[11\={2}]}

\newcommand {\rcut}{E_{\rm cut}}
\newcommand {\Ecoh}{E_{\rm coh}}
\newcommand {\dyield}{d_{\rm yield}}

\begin{document}

\date{\today
 }

\title{Effect of material stiffness on hardness: a computational study
based on model potentials }

\author{Gerolf Ziegenhain}
\affiliation{%
Fachbereich Physik, Universit{\"a}t Kaiserslautern, \\
Erwin-Schr{\"o}dinger-Stra{\ss}e, D-67663 Kaiserslautern, Germany}

 \author{Herbert M.~Urbassek}
 \email{urbassek@rhrk.uni-kl.de}
 \homepage{ http://www.physik.uni-kl.de/urbassek/}
\affiliation{%
Fachbereich Physik, Universit{\"a}t Kaiserslautern, \\
Erwin-Schr{\"o}dinger-Stra{\ss}e, D-67663 Kaiserslautern, Germany}

\begin{abstract}

We investigate the dependence of the hardness of materials on their
elastic stiffness. This is possible by constructing a series of model
potentials of the Morse type; starting on modelling natural Cu, the
model potential exhibit an increased elastic modulus, while keeping all
other potential parameters (lattice constant, bond energy) unchanged.
Using molecular-dynamics simulation, we perform nanoindentation
experiments on these model crystals. We find that the crystal hardness
scales with the elastic stiffness. Also the load drop, which is
experienced when plasticity sets in, increases in proportion to the \es,
while the yield point, i.e., the indentation at which plasticity sets
in, is independent of the \es.

\end{abstract}


\maketitle


\section{Introduction}

While the elastic properties of solids are well understood, and their
description in terms of interatomic potentials is well established,
plastic deformation of solids -- and the concurrent phenomena of
dislocation generation and motion -- are more complex and their
modelling presents greater difficulties. One reason certainly is that
plasticity involves strongly non-equilibrium states in the solid, in
which the material is stressed so far that interatomic bonds are broken
and new bonds are formed.

The simplest quantitative measure of plasticity is given by the material
hardness, i.e., the pressure with which the material withstands plastic
deformation. It has long been known that the hardness of a defect-free,
ideal crystal, the theoretical strength, is proportional to the
material's shear modulus. This important result, which dates back to
Frenkel,\cite{Fre26,KM87b} has been derived by considering the shear
stress necessary to induce slip in a perfect lattice; modern ab initio
quantum mechanical calculations have confirmed this
result.\cite{RKCM99,OLY02,OLH*04} However, the question remains how the
\es\ of a material influences dislocation generation and the emergence
of plasticity in a more complex and realistic situation.

We choose a nanoindentation scenario to investigate the onset of
plasticity, and its dependence on the material stiffness. A
molecular-dynamics simulation allows to provide atomistic insight into
the reaction of a material to an applied load, to calculate the
force-depth curve and to extract the contact pressure and the material
hardness. It furthermore allows to describe in detail the induced
damaged patterns in the form of dislocation loops and stacking faults
formed.

Our study is based on the Morse interatomic potential. While it is well
known that metals cannot be described in all details by pair
potentials,\cite{Car90} this class of potentials readily allows to
generate a series of potentials, which all describe the same material,
in which, however, exactly one property is arbitrarily changed. Since
the prime aim of our study is to inquire into the generic dependence of
plasticity and hardness on the \es\ of the material, rather than to
describe one particular material as accurately as possible, our choice
of a Morse potential appears appropriate. We note that the Morse
potential has been used previously to describe dislocations in
metals.\cite{CD66,Vit68,Vit88,HMP89}

\section{Method}

\subsection{Potentials}

We use the Morse potential

\begin{equation}  \label{pot}
V(r)=D\left\{\exp\left[-2\alpha(r-r_0)\right]-2\exp\left[-\alpha
(r-r_0)\right]\right\}
\end{equation}

to model our material. It is characterized by three parameters: the bond
strength $D$, the equilibrium bond distance $r_0$, and the potential
fall-off $\alpha$. As is well known,\cite{Car90,Fin03} the Morse
potential cannot describe all characteristics of bonding in metals;
however, as discussed elsewhere,\cite{ZHU} it can give a reasonable
description of the elastic\cite{GW59,LKG67} and also plastic processes
occurring under indentation. We adopted this potential, since it allows
to fit in a transparent way the materials properties. Since the Morse
potential contains three parameters, it is possible to fit it to three
materials properties; these are traditionally chosen as the lattice
constant $a$, the cohesive energy $\Ecoh$, and the bulk modulus $B$.

For example, Cu\cite{Smi49} with $a=3.615$ \AA, $\Ecoh=3.54$ eV, and $B=
134.4$ GPa can be described by a Morse potential with $D= 0.337$ eV,
$r_0 = 2.89$ \AA, and $\alpha= 1.33$ \AA$^{-1}$. We note that here and
in the following, we cut off the potential at $\rcut= 2.5a = 9.0375$
\AA\ -- i.e. including 248 neighbours -- , and shift the potential to
zero at this distance.

We create a series of Morse potentials, in which the lattice constant
and cohesive energy are kept unchanged, but the bulk modulus is set to a
preassigned value. These potentials can hence be considered as
describing a series of \emph{pseudo-Cu} materials with identical bond
strength, but changed bulk moduli. Table 1 reproduces the values of the
fit parameters obtained for the potentials employed in this study. We
use a Levenberg-Marquardt based optimization for parameter fitting; the
values of $a$ ($\Ecoh$) are reproduced within 0.1 \% (1\%).

\begin{figure}

   \includegraphics[width=11cm]{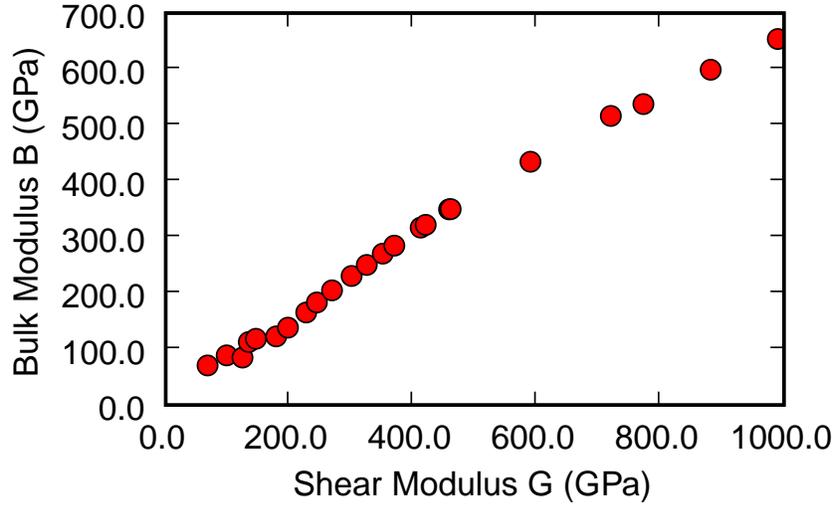}

\caption{ Correlation between the shear modulus $G$ and the bulk modulus
$B$ in the series of Morse potentials investigated.
 }

\label{f_GB}
\end{figure}

The linear elastic properties of an fcc crystal are given by 3 elastic
constants, $c_{11}$, $c_{12}$, and $c_{44}$. However, for a pair
potential, it is always $c_{12}$= $c_{44}$, so that only two elastic
constants describe the elastic behaviour. We chose the bulk modulus

\begin{equation}  \label{B}
B = \frac{c_{11}+ 2  c_{12}}{3}
\end{equation}

and the average shear modulus

\begin{equation}  \label{G}
G = \frac{ c_{11}+ 2 c_{44}- c_{12} }{5}
\end{equation}

to describe the elastic properties. For Cu, it is $G= 39.8$ GPa. Table 1
shows the shear moduli obtained for pseudo-Cu, and \qfig{f_GB}
demonstrates that $G$ increases quite linearly with $B$; only for the
highest moduli, $G$ increases superlinearly with $B$. So in general, we
may say that the entire elastic behaviour (the \es) of pseudo-Cu changes
in correspondence with $B$; we shall talk of \emph{weak} and
\emph{strong} materials.

\subsection{Generalized stacking fault energy}

The generalized stacking fault (GSF) energy can be used to characterize
the behaviour of a material with respect to formation of stacking
faults, and hence dislocation formation and
glide.\cite{Vit68,Ric92,FZV*97,LKBK00} For its definition, we consider
an fcc crystal, whose upper part has glided along a (111) plane with
respect to its lower part by a definite amount; in the present paper we
only consider glide vectors along the \two\ direction. The GSF energy is
the potential energy per surface area of the deformed crystal; when
determining this energy, relaxation of the crystal vertical to the (111)
plane, but no relaxation or reconstruction within this plane is allowed
for.\cite{Vit68,FZV*97} In our calculation, it proved necessary to
choose the crystallite rather large, 10 lattice in lateral directions
and 25 lattice constants in both vertical directions; we employed the
conjugate-gradient technique to relax the crystal in vertical direction.

\begin{figure}

   \includegraphics[width=11cm]{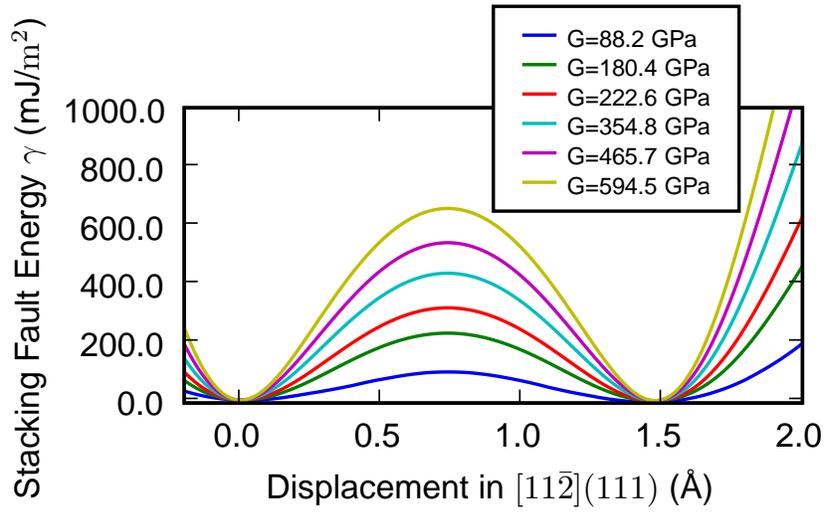}

\caption{Generalized stacking fault energy as a function of displacement
in \two\ direction in the series of Morse potentials investigated.
 }

\label{f_gsf}
\end{figure}

\qfig{f_gsf} displays the variation of the GSF surface along the \two\
(111) displacement. The minimum at 0 displacement corresponds to fcc
stacking, here $\gamma=0$ by definition. The second minimum is found for
a displacement corresponding to a partial Burgers vector $\frac{1}{6}
[\bar{1}\bar{1}2] a = a/\sqrt{6} \cong 1.5$ \AA. It describes a stacking
fault with energy $\gams$. For our potentials, extremely small values
were obtained, which varied in the range of $\gams = (-2 \dots +2)$
mJ/m$^2$. For comparison, the experimental value of $\gams$ for Cu is 45
mJ/m$^2$; \cite{HL82,MMP*01} this value is retrieved by ab initio
calculations.\cite{RKCM99,OLY02} We note that in previous
investigations, a sufficiently large value of the stacking fault energy
could only be obtained by adapting the value of the cut-off radius to
$\rcut =2.2a$;\cite{CD66} for larger $\rcut$, $\gams$ strongly
decreased.

We found that the exact determination of $\gams$ is nontrivial, since
large crystallites have to be relaxed; sometimes only a local and not
the global energy minimum may have been found. Furthermore, it is known
that the stacking fault energy may depend sensitively on the cut-off
radius of the potential.\cite{CD66} Hence we conclude that within the
Morse pair potential approximation, the stable stacking fault energy of
fcc metals (of Cu, at least) is strongly underestimated. It does not --
or only negligibly -- depend on the elastic stiffness of the material.

\begin{figure}

   \includegraphics[width=8cm]{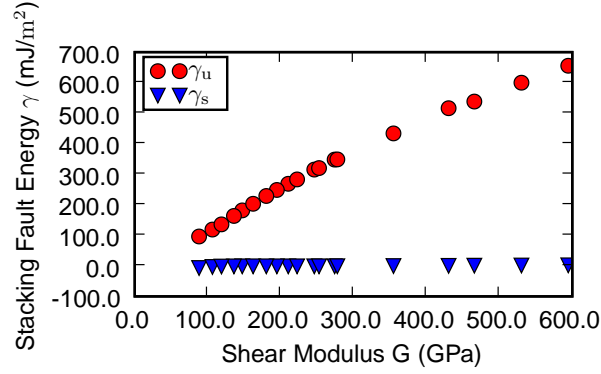}

\caption{Dependence of the stable and unstable \SF\ energies, $\gamu$
and $\gams$ on the shear modulus $G$ in the series of Morse potentials
investigated.
 }

\label{f_sfB}
\end{figure}

The energy barrier between the fcc crystal position and the stable
stacking fault is called the unstable stacking fault energy, $\gamu$. It
depends strongly on the elastic stiffness. As \qfig{f_sfB} shows, it
increases roughly linearly with the bulk modulus $B$, and hence also
with the shear modulus $G$, cf.\ \qfig{f_GB}. This may be understood
since the displacement of two (111) planes in the crystal corresponds to
a gliding motion; thus the barrier to gliding, $\gamu$, should be
strongly correlated with the shear modulus.

\subsection{Indentation}

In order to model the plastic deformation of our material, we performed
indentation simulations using the a appropriately adapted version of the
LAMMPS molecular dynamics code.\cite{LAMMPS} Our target consists of an
fcc crystallite with a (100) surface; it has a depth of 25 nm and a
square surface area of 621.2 nm$^2$; it contains 1325598 atoms.

The indenter is modelled as a soft sphere. We chose a non-atomistic
representation of the indenter, since we are not interested in the
present study in any atomistic displacement processes occurring in the
indenter, but only in the substrate. The interaction potential between
the indenter and the substrate atoms is characterized by a hard core of
radius $R$ surrounded by a softly repulsive potential\cite{KPH98}

\begin{equation} \label{5}
V(r) = \left\{ \begin{array}{ll}
k(R-r)^3, & r<R, \\
0, & r\ge R .
\end{array} \right.
\end{equation}

since we are in this generic study not interested in the complexities
introduced by adghesion phenomena. The indenter stiffness was set to
$k=3$ \eVA. Our indenter has a radius of $R=8$ nm. Indentation proceeds
in the so-called \emph{velocity-controlled} approach, in which the
indenter proceeds with a fixed velocity, $v=20$ m/s in our case, into
the substrate.

\section{Results} \label{Results}

\subsection{Force-displacement curves}

Fig.~4 shows the basic result of the simulation, the force-displacement
curves. In all these curves it is seen that the force $F$ increases
monotonically with the displacement $d$ into the substrate until a depth
$\dyield$ which is of the order of $9.36\pm 0.48$ \AA\ and where the
force suddenly drops. This corresponds to the onset of plastic
deformation inside the material. The first, monotonically increasing
part is due to elastic deformation of the substrate. Hertz\cite{Her82}
calculated that for an elastically isotropic solid it holds

\begin{equation} \label{Hertz}
 F = \frac43 \Er d^{3/2} \sqrt{R} .
\end{equation}

In this relation, a single materials parameter, the so-called reduced
modulus $\Er$ describes the materials elastic
response.\cite{Fis04,Fis07} For a rigid indenter, it may be expressed in
terms of the Young's modulus $E$ and the Poisson ratio $\nu$ of the
substrate as

\begin{equation} \label{Er}
\Er = \frac{E}{1-\nu^2} .
\end{equation}

Hertz also determined the contact pressure $p$; it is defined by the
ratio of the normal force $F$ divided by the contact area projected into
the surface plane, $A$. We calculate $A$ from the ensemble of contact
atoms of the indenter, which are farthest away from the indentation
axis, and approximating it as an ellipse.

We test this law in Fig.~4b. To do this, we need to calculate the
reduced elastic modulus for a (100) surface. Here, Young's modulus has
to be calculated for deformation in (100) direction, perpendicular to
the surface, while the Poisson ratio has to be averaged over all
directions perpendicular to the (100) directions. The corresponding
formulae are rather complex and are provided in \qref{TS71}. Fig.~4b
shows that in the range of $G= 80 - 180$ GPa, the elastic behaviour
follows quite uniquely the (generalized) Hertz law \queq{Hertz}. Only
for the extreme cases of $G= 39$ and 278 GPa, the normalized forces
$F/\Er$ are too small. We believe that this deviation from the Hertz law
is due to (i) a poorer quality of the potential to describe the
materials reaction (ii) in particular in the case of the weaker
pseudo-Cu, $G=39$ GPa, the softness of the substrate makes the
fluctuations in the response of the substrate to the constant-velocity
indentation more sizable. We note that when normalizing $F/G$, an
equally satisfactory uniformity of the curves is achieved; we do not
show this plot, since the theoretical foundation appears to be missing.

At the yield point the force drops suddenly due to the onset of
plasticity; this well-known phenomenon is called the \emph{load drop}.
In our series of simulations we saw that the exact positions of the
yield point $\dyield$ does not show a monotonic trend with the \es\ of
the substrate; rather $\dyield$ fluctuates. We believe that in this
constant-velocity indentation, the yield point may be subject to
fluctuations in the simulation procedure. We therefore conclude that
within the limits of the fluctuations the position of the yield point
does not depend on the \es\ of the target.

However, the load drop shows a clear increase with the \es\ of the
material. This is understandable since the atomistic reason for the load
drop is the nucleation of a stacking fault in the material. Its
generation requires an energy which scales with the unstable stacking
fault energy, $\gamu$, which has to be found to scale well with the \es.

\subsection{Hardness}

The force $F$ divided by the (projected) contact area $A$ of the
indenter defines the contact pressure $p$. After the onset of
plasticity, it gives the hardness of the material. Fig.~5 shows the
simulation results for the contact pressure. In view of the good quality
of the scaling observed before with the material stiffness, we present
our data scaled with the shear modulus $G$. This is also motivated by
the consideration that the theoretical shear strength $\tau$ of the
material is given by

\begin{equation} \label{tau}
\tau = \eps G   ,
\end{equation}

where $\eps$ is a constant, which depends on the crystal structure of
the solid but is otherwise quite material-independent. Frenkel estimated
$\eps = 1/(2\sqrt{2}\pi) = 0.11$ for the important $\langle \bar{1}
\bar{1} 2 \rangle$ \{111\} glide system of fcc solids.\cite{KM87b}
Recent more refined estimates based on density functional theory give a
slightly reduced value, $\eps = 0.085$.\cite{RKCM99,OLY02,OLH*04}

Tabor showed that the hardness measured as the contact pressure during
nanoindentation amounts to $3\tau$,\cite{Tab51,KM87b} since the
nanoindentation acts as a `lens´ focussing the stress in a small volume
beneath the contact point.\cite{Li07} Using \qeq{tau}, we hence expect

\begin{equation} \label{H}
H= 3\tau = 3\eps G  .
\end{equation}

Fig.~5 shows that after the onset of plasticity, the hardness is indeed
proportional to the shear modulus and well described by \qeq{H} with a
coefficient of $0.2- 0.25$, i.e., $\eps= 0.07 - 0.08$, in satisfactory
agreement with the estimates of $\eps$ given above.. The figure also
appears to indicate that the fluctuations in the fully plastically
regime are smaller for soft materials, which appears plausible.

Note that also in the elastic regime, the scaled contact pressure $p/G$
lies on one single curve, again underlying the importance of the shear
modulus for the indentation behaviour. The only exception is given by
the softest material.

In this case the energy difference between the fcc and the hcp
structures is rather small, and we observe instabilities at the relaxed
free surface of the crystallite, which are reflected in the hardness
curve, Fig.~5.

\subsection{Plasticity}

Figs,~6 and 7 give an atomistic presentation of the dislocations which
developed in the substrate after the onset of plasticity. We analyzed
the local atomic structures using a method based on the angular
correlation of nearest-neighbour atoms.\cite{AJ06} We found this method
to be superior for our purposes than the more common analysis based on
the centrosymmetry parameter.\cite{KPH98} Only atoms deviating from the
fcc structure are visualized: the hcp stacking faults (red) are
surrounded by unidentifiable structures of low symmetry (grey) and a
very small amount of particles in a bcc coordination (green). These grey
and green atoms occur at the boundaries of the dislocation loops due to
the strong lattice deformations existing there and mark the ends of the
loops. The free surface is also coloured grey.

Fig.~6 displays the plasticity shortly after the nucleation of the first
dislocations; note that the two materials shown here have a similar
yield point (cf.\ Fig.~4), so that the amount of dislocated material can
be compared. The damage is more concentrated around the indenting sphere
for the stiffer material. Note also that the size of the dislocation
loops has increased for the softer material.

For fully developed plasticity (Fig.~7) we observe how prismatic
dislocation loops have formed and were driven away from the indenter. In
all cases which we studied, the $\langle \bar{1} \bar{1} 2 \rangle$
\{111\} glide system was activated first, and later, i.e. under larger
stress gradients, also the $\langle 1 \bar{1} 0 \rangle$ \{111\} glide
system. This is a mere consequence of crystallography and is reflected
by the smaller Schmid factor for the latter glide system; as expected,
the changed \es\ in our model crystals does not influence which glide
systems are activated. However, it does affect the form and size of the
plastic zones. Thus, for smaller \es, (Fig.~7a), evidently \emph{more}
and \emph{smaller} loops have been generated. This is in agreement with
our finding that for smaller stiffness, also the unstable stacking fault
energy -- and hence the barrier to dislocation formation and slip -- is
smaller: hence we have \emph{more} loops. In contrast, for the stiffer
material (Fig.~7b), the nucleation of loops is retarded, and fewer but
\emph{larger} loops are formed.

\section{Conclusions}

\begin{enumerate}

\item The Morse potential allows to systematically change the materials
properties. We concentrate on changing the bulk modulus (and thus the
elastic stiffness), while keeping the cohesive energy (bond strength)
constant. In this potential, all elastic constants are changed in
proportion to each other.

\item Also the unstable stacking fault energy, which can be viewed as
the resistance of a dislocation to gliding, changes in proportion to the
elastic stiffness. The stable stacking fault energy, which is relevant
for the width of partial dislocations, is quite unaffected from changes
in the \es.

\item Not unexpectedly, the elastic part of the indentation curve scales
with the \es.

\item The yield point, i.e., the indentation at which plasticity sets
in, does not depend -- apart from fluctuations -- on the \es.

\item The load drop, which is experienced when plasticity sets in,
increases in proportion to the \es.

\item The material hardness $H$ is proportional to the theoretical
strength, and thus on the \es; it is $H= (0.20 - 0.25) \cdot G$.

\end{enumerate}

\begin{acknowledgments}

The authors acknowledge financial support by the Deutsche
Forschungsgemeinschaft via the Graduiertenkolleg 814.

\end{acknowledgments}


\bibliography{c:/D/bib/base/string,c:/D/bib/base/all,c:/D/bib/base/publ}


\begin{table}

   \begin{tabular}{lll|ll|l}

$D$ (eV) & $\alpha$ (\AA$^{-1}$) & $r_0$ (\AA) & $B$ (GPa) & $G$ (GPa) & $\gamma_u$ (mJ/m$^2$) \\ \hline

 0.203 &0.90   &3.47 & 67.7       &39.2    &71.8          \\
 0.272 &1.12   &3.10 &  98.6      &58.6    &89.9           \\
 0.320 &1.28   &2.93 &  124.2     &74.4    &85.8           \\
 0.337 &1.33   &2.89 & 134.4      &80.5    &113.8         \\
 0.359 &1.41   &2.83 &  145.9     &88.2    &96.2          \\
 0.390 &1.54   &2.77 &  178.7     &106.5   &118.9     \\
 0.416 &1.65   &2.72 & 197.7      &118.9   &135.9     \\
 0.437 &1.76   &2.69 &  227.3     &136.0   &163.5     \\
 0.455 &1.86   &2.66 &  244.5     &147.4   &181.6     \\
 0.471 &1.96   &2.64 &  269.0     &162.3   &203.3     \\
 0.484 &2.05   &2.63 & 300.7      &180.4   &228.9     \\
 0.495 &2.13   &2.62 &  325.2     &194.9   &248.6     \\
 0.504 &2.22   &2.61 &  351.1     &210.5   &268.8     \\
 0.513 &2.30   &2.60 &  369.8     &222.6   &283.1    \\
 0.520 &2.38   &2.60 &  412.4     &246.2   &315.3     \\
 0.526 &2.45   &2.59 &  420.5     &252.7   &320.2     \\
 0.532 &2.52   &2.59 &  458.3     &274.1   &348.0    \\
 0.537 &2.59   &2.58 & 460.9      &277.8   &348.6    \\
 0.555 &2.92   &2.57 &  590.0     &354.8   &433.8     \\
 0.563 &3.16   &2.57 &  720.0     &430.4   &516.3     \\
 0.571 &3.38   &2.56 &  772.7     &465.7   &538.1     \\
 0.589 &3.54   &2.56 &  881.5     &530.3   &599.9     \\
 0.591 &3.73   &2.56 &  990.1     &594.5   &655.3     \\
   \end{tabular}

\caption{Fitted parameters of the potentials, $D$, $\alpha$, $r_0$, cf.\
\qeq{pot}. Materials properties determined from these potentials, $B$,
$G$, $\gamu$. }

   \label{t_potentials}
\label{t1} \end{table}


\begin{figure}

\subfigure[]{\includegraphics[width=8cm]{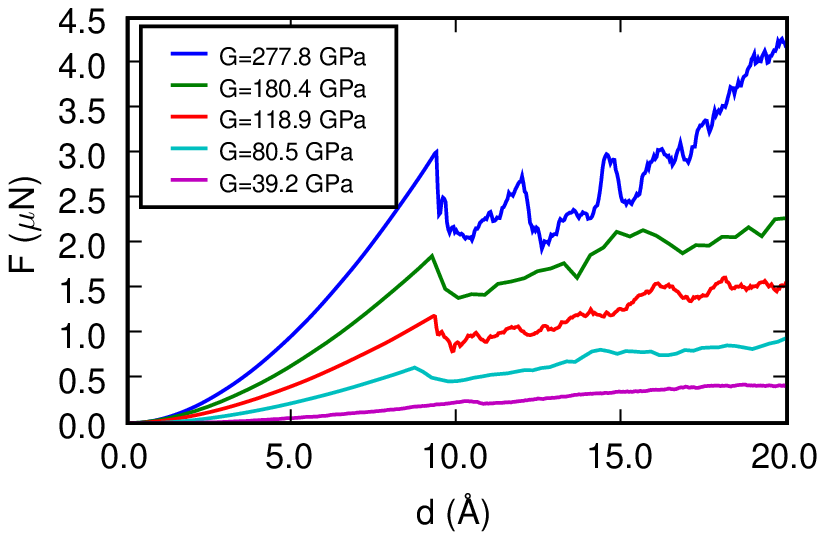}} \hfill
\subfigure[]{\includegraphics[width=8cm]{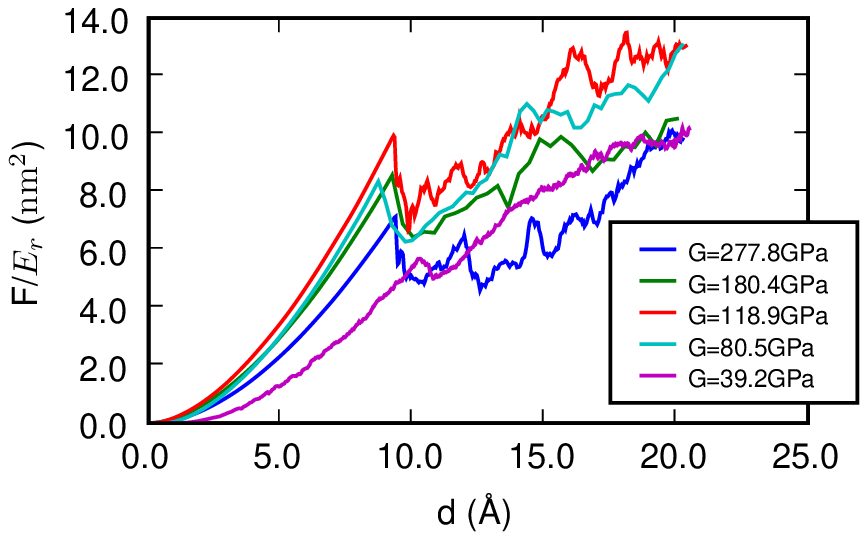}}

\caption{a) Dependence of the indentation force $F$ on the indentation
depth $d$ in the series of Morse potentials investigated. b) Forces $F$
normalized to the reduced elastic modulus $\Er$.
 }

\label{f_Fd}
\end{figure}

\begin{figure}

   \includegraphics[width=8cm]{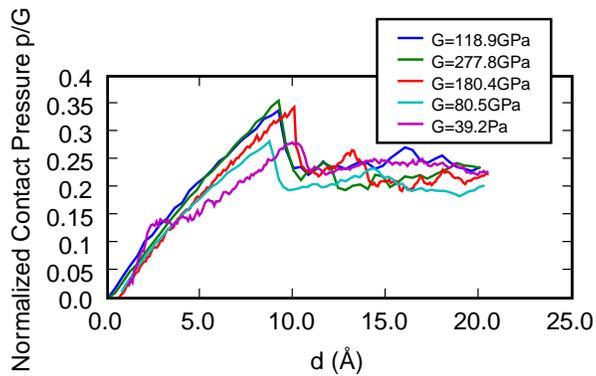}

\caption{Contact pressure $p$, normalized to the shear modulus $G$, as a
function of indentation depth $d$ in the series of Morse potentials
investigated. After the onset of plasticity, the contact pressure
defines the hardness of the material.
 }

\label{f_Hd}
\end{figure}

\begin{figure}

\subfigure[]{\includegraphics[width=0.65\textwidth]{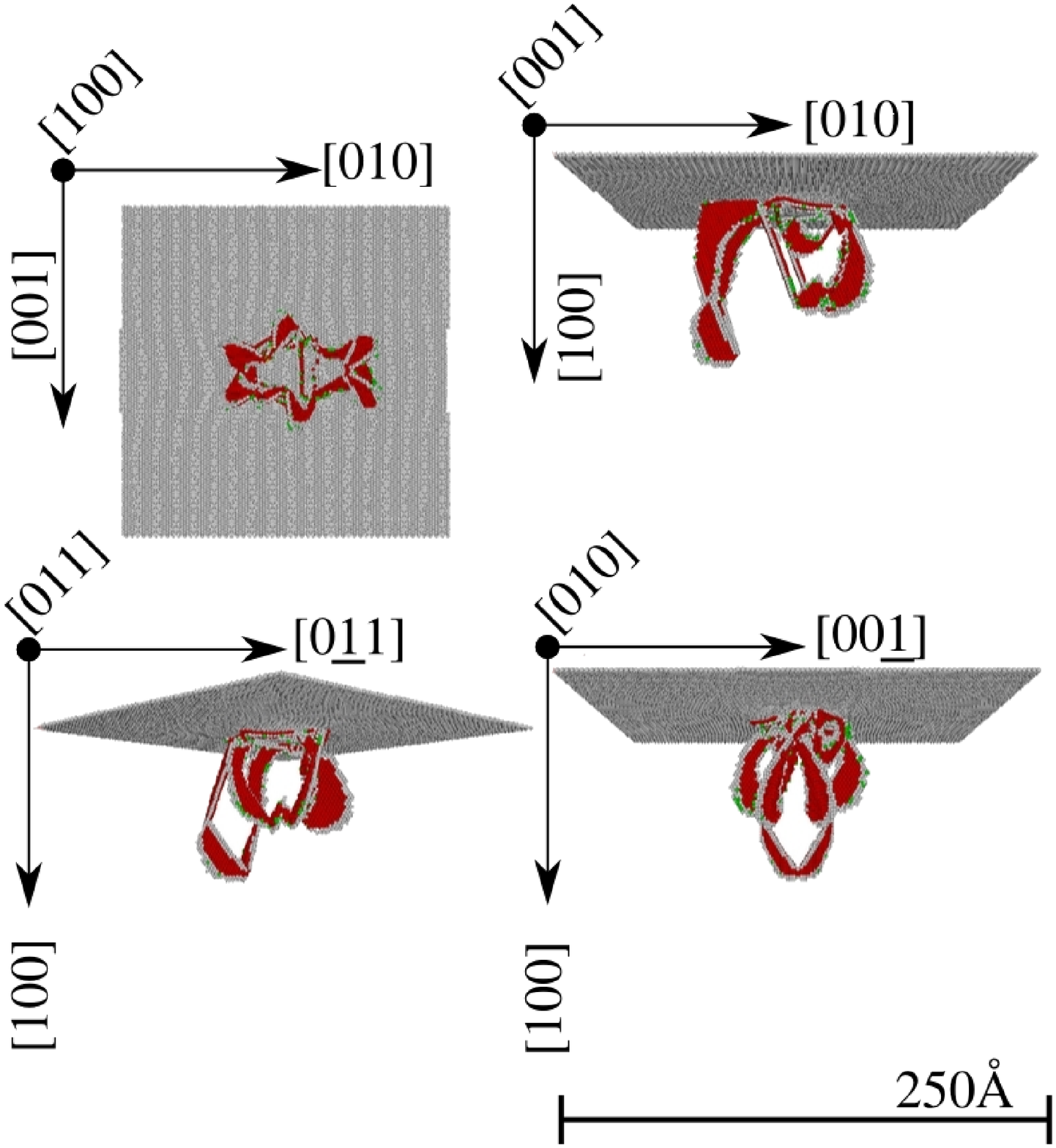}}
\\
\subfigure[]{\includegraphics[width=0.65\textwidth]{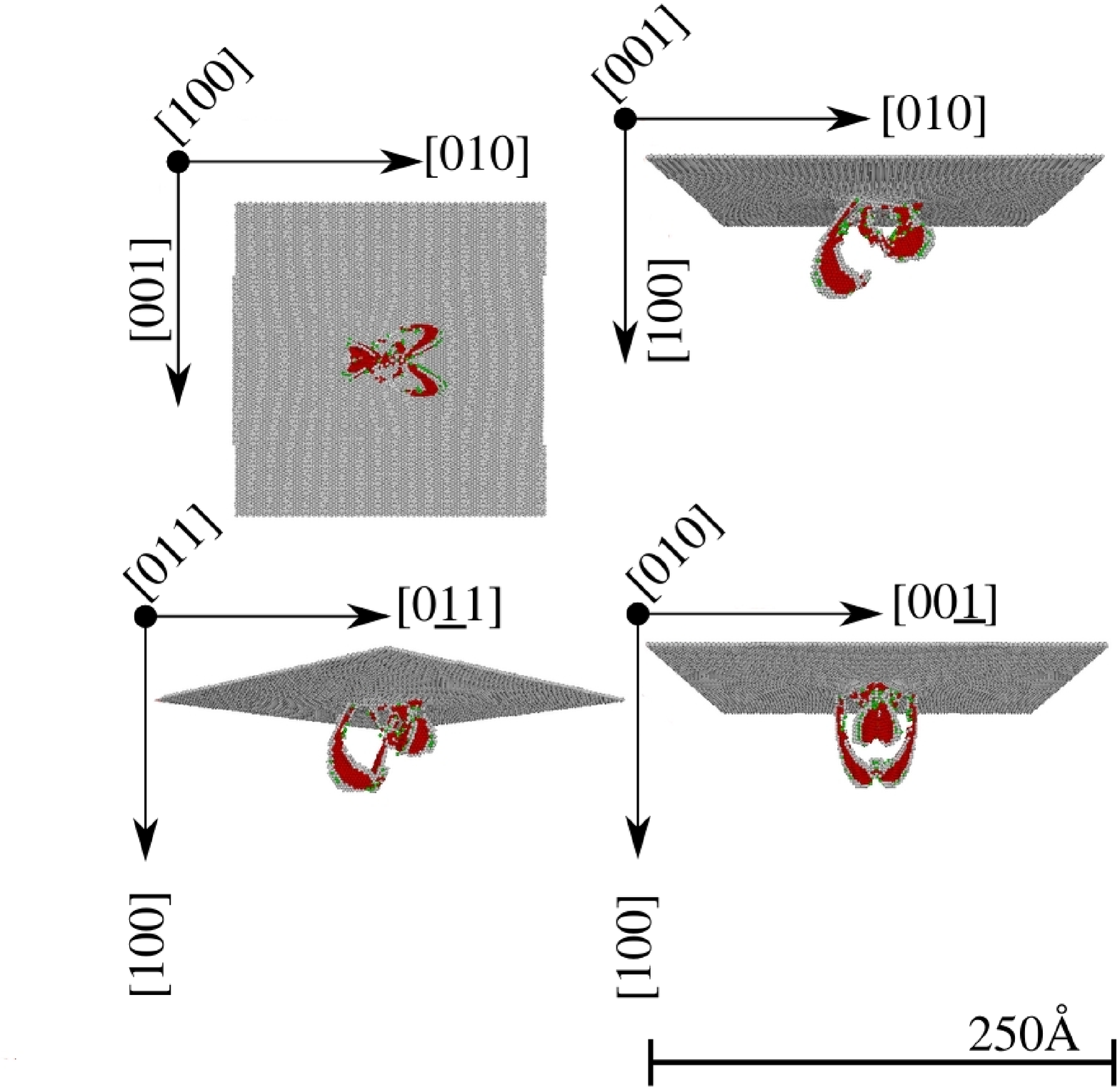}}

\caption{Incipient plasticity determined at an indentation depth of
$d=9.7$ \AA\ for an indentation target with bulk modulus of (a)
$B=134.4$ GPa b) $B=300.7$ GPa. In each subfigure, the plots show a view
from the bottom and three perspective side views, which can be
identified by the axes drawn. The grey area shows the surface. The atoms
colored in red display dislocation loops.
 }

\label{f_inc}
\end{figure}

\begin{figure}

\subfigure[]{\includegraphics[width=0.7\textwidth]{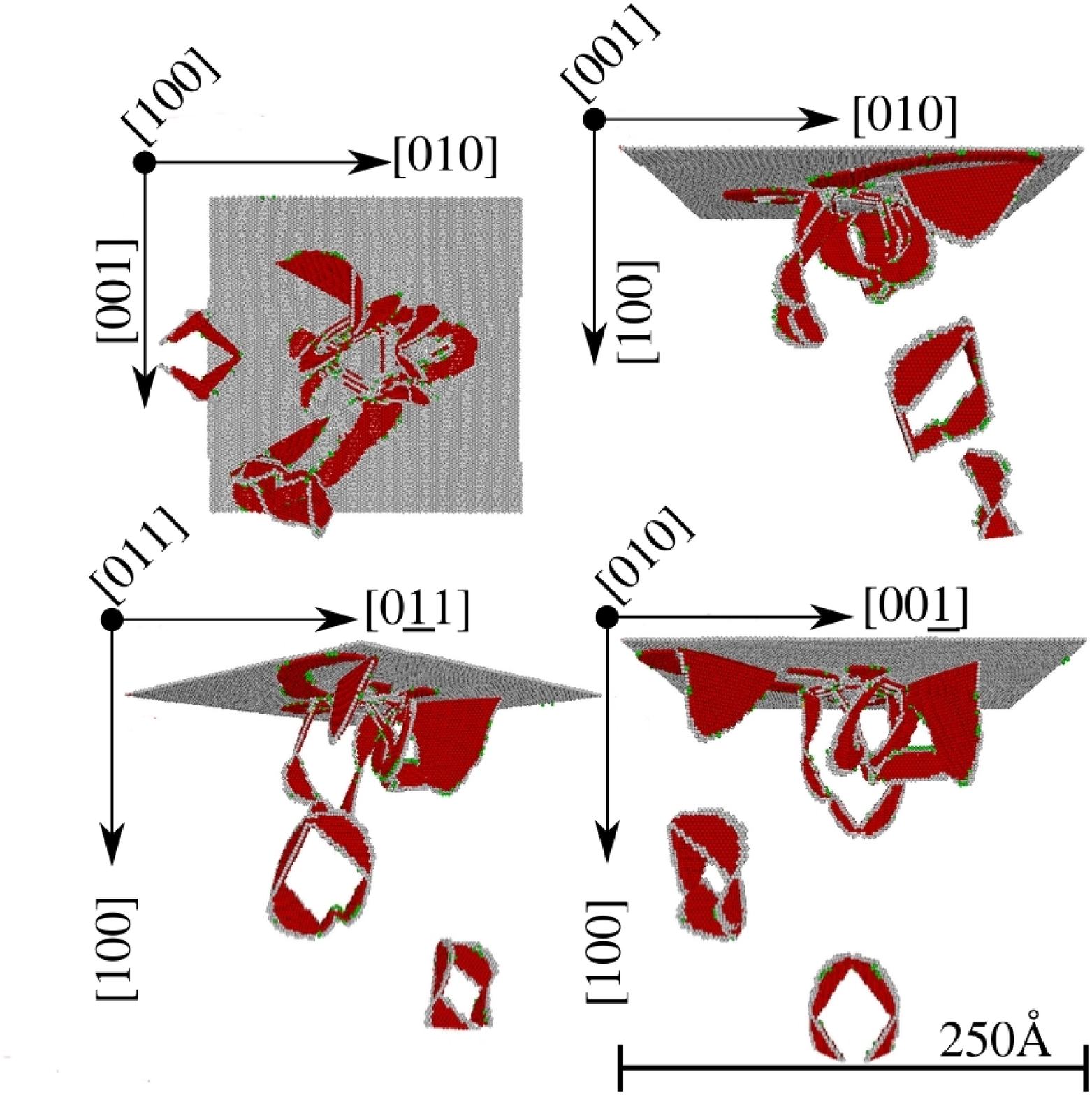}}  \\
\subfigure[]{\includegraphics[width=0.7\textwidth]{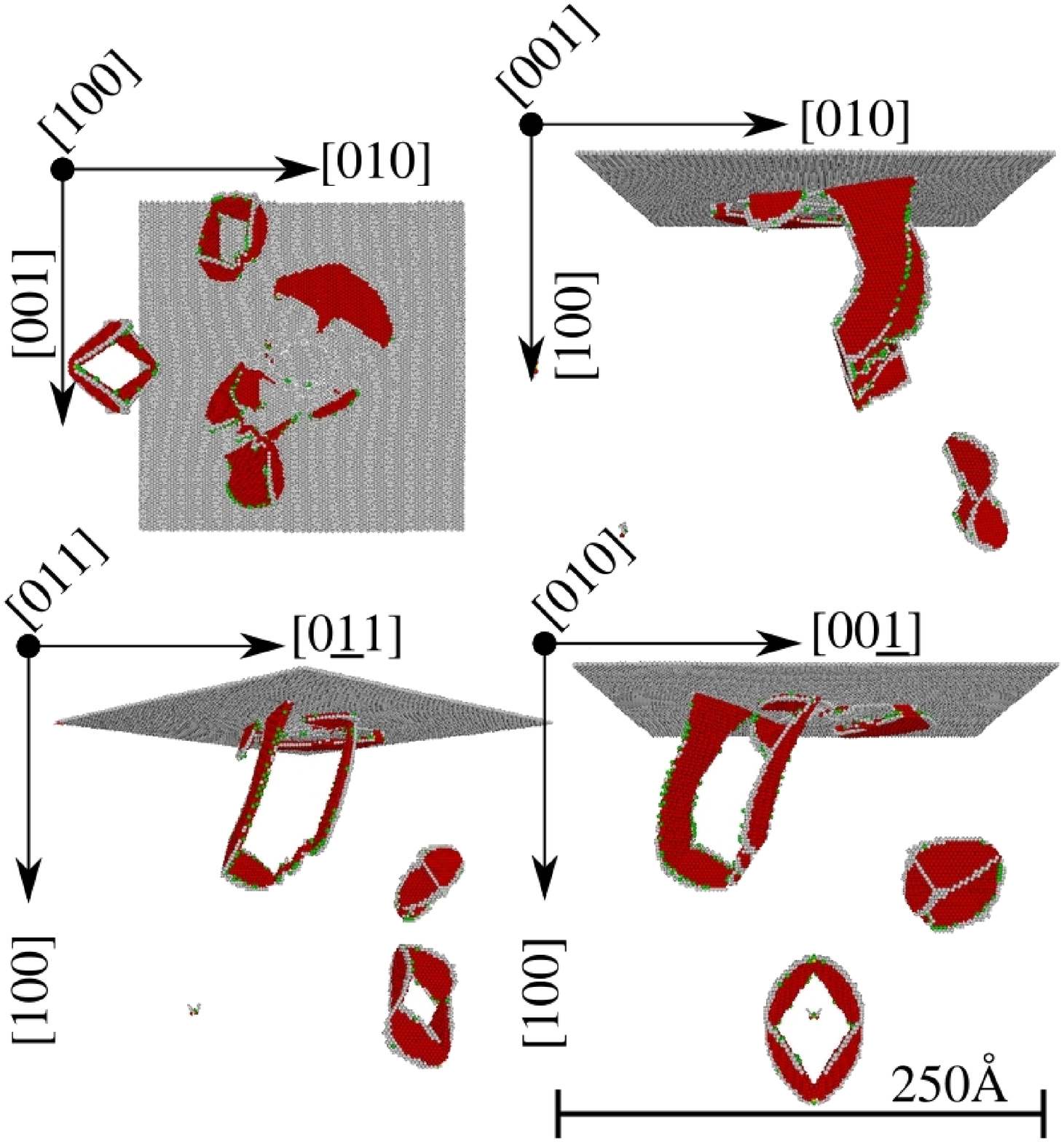}}

\caption{Fully developed plasticity determined at an indentation depth
of $d=11.9$ \AA\ for the two cases shown in \qfig{f_inc}.
 }

\label{f_full}
\end{figure}

\clearpage

\end{document}